\documentclass[12pt]{article}
\pdfoutput=1
\usepackage[utf8]{inputenc}
\usepackage{amsmath}
\usepackage{amssymb}
\usepackage{units}
\usepackage{graphicx}
\usepackage{slashed}
\usepackage{array}
\usepackage[leftcaption,raggedright]{sidecap}
\usepackage{caption}
\usepackage[hidelinks]{hyperref}

\newcommand\pubnumber{NuPhys2026-Avelino-Vicente}
\newcommand\pubdate{\today}

\def\IFIC{
Instituto de F\'{i}sica Corpuscular, \\
  CSIC-Universitat de Val\`{e}ncia, 46980 Paterna, Spain}

\def\support{\footnote{
  Work supported by the Spanish grants PID2023-147306NB-I00, CNS2024-154524 and CEX2023-001292-S (MICIU/AEI/10.13039/501100011033), as well as CIPROM/2021/054 (Generalitat Valenciana).
}}

\def\Title#1{\begin{center} {\Large #1 } \end{center}}
\def\Subtitle#1{\begin{center} {\large #1 } \end{center}}
\def\Author#1{\begin{center}{ \sc #1} \end{center}}
\def\Address#1{\begin{center}{ \it #1} \end{center}}

\newcommand\pubblock{\rightline{\begin{tabular}{l} \pubnumber\\
         \pubdate  \end{tabular}}}
\newenvironment{Abstract}{\begin{quotation}  }{\end{quotation}}
\newenvironment{Presented}{\begin{quotation} \begin{center} 
             PRESENTED AT\end{center}\bigskip 
      \begin{center}\begin{large}}{\end{large}\end{center} \end{quotation}}

\def\beq{\begin{equation}}
\def\eeq#1{\label{#1}\end{equation}}
\def\eeqn{\end{equation}}

\def\beqa{\begin{eqnarray}}
\def\eeqa#1{\label{#1}\end{eqnarray}}
\def\eeqan{\end{eqnarray}}

\let\bar=\overbar

\def\Dslash{\not{\hbox{\kern-4pt $D$}}}
\def\dslash{\not{\hbox{\kern-2pt $\del$}}}

\def\msb{{\bar{\ssstyle M \kern -1pt S}}}

\begin{document}
\begin{titlepage}
\pubblock

\vfill
\Title{Neutrino mass models}
\Subtitle{A short review with emphasis on the majoron}
\vfill
\Author{Avelino Vicente\support}
\Address{\IFIC}
\vfill
\begin{Abstract}
Neutrino masses provide one of the clearest indications of physics beyond the Standard Model. In this brief review, I discuss the main theoretical frameworks developed to account for them, with particular emphasis on scenarios in which neutrinos are Majorana particles. After a short overview of the current landscape of neutrino mass models, I focus on constructions featuring the spontaneous breaking of global lepton number and examine the phenomenological implications of a massless Goldstone boson, the majoron.
\end{Abstract}
\vfill
\begin{Presented}
NuPhys2026, Prospects in Neutrino Physics\\
King's College, London, UK,\\ January 7--9, 2026
\end{Presented}
\vfill
\end{titlepage}
\def\thefootnote{\fnsymbol{footnote}}
\setcounter{footnote}{0}

\section{Introduction}
\label{sec:intro}

The Standard Model (SM) of particle physics stands as one of the most impressive human achievements, providing an excellent description of a wide range of experimental observations. However, it is not the final theory. Several theoretical issues and experimental problems call for an extension. Among them, the existence of non-zero neutrino masses and lepton mixing~\cite{deSalas:2020pgw} is arguably the clearest.

At present, there are several open questions in neutrino physics. Among them:

\begin{enumerate}
\item What is the origin of neutrino masses?
\item Are neutrinos Dirac or Majorana fermions?
\item What is the absolute scale of neutrino masses?
\item What is the mass ordering?
\item Are there more than three neutrinos? Maybe sterile?
\item Is there CP violation in the lepton sector?
\end{enumerate}

In this short review I will mainly concentrate on questions 1 and 2. While the answer to all questions can only be obtained experimentally, in the case of these two questions theory input, in the form of model building, is also required. 

The rest of the paper is organized as follows: In Sec.~\ref{sec:general} I discuss several general facts on neutrino mass models, including the crucial distinction between Dirac and Majorana mass models. Then, in Sec.~\ref{sec:majoron} I concentrate on Majorana neutrino mass models with spontaneous breaking of a global lepton number symmetry. These models include a massless Goldstone boson, the majoron, with many interesting phenomenological implications. Some final remarks are given in Sec.~\ref{sec:final}.

\section{Generalities}
\label{sec:general}

Many extensions of the SM that incorporate a mechanism to generate neutrino masses have been proposed over the years. These proposals vary in several respects. Some induce neutrino masses at tree level, while others do so via radiative corrections; some involve new degrees of freedom at the TeV scale or below, while others introduce very heavy states; and some focus solely on neutrino masses, while others also address additional open issues. In these models (at least in those that are theoretically appealing), not only do neutrinos acquire masses, but their smallness is also explained. See~\cite{Boucenna:2014zba,King:2017guk,Cai:2017jrq,Avila:2025qsc} for some reviews.

From a model-building point of view, the first and foremost question is whether neutrinos are Dirac or Majorana fermions. This distinction is crucial, as the entire setup depends on it. In the former case, \textbf{Dirac neutrinos}, all SM fermions would share the same nature and lepton number could be a conserved symmetry. The latter case, \textbf{Majorana neutrinos}, is more economical in terms of degrees of freedom, but introduces a novel type of fermion and entails lepton number violation.

The \textit{Dirac vs.\ Majorana question} can only be settled by observing a lepton-number-violating process. For instance, the observation of neutrinoless double beta decay would imply that neutrinos are Majorana particles, as indicated by the well-known Schechter–Valle black-box theorem~\cite{Schechter:1981bd}. In the meantime, and in the absence of experimental hints favoring either possibility, there is room for speculation in both directions.

\subsection{Dirac neutrinos}
\label{subsec:dirac}

\begin{table}
  \renewcommand*{\arraystretch}{1.4}
  \centering
  \begin{tabular}{|ccccc|}
    \hline
    & $SU(2)_L$ & $U(1)_Y$ & $U(1)_L$ & $\mathbb{Z}_2$ \\
    \hline
    \hline
    $\nu_R$ & $\mathbf{1}$ & $0$ & $1$ & $+$ \\
    $N_{L,R}$ & $\mathbf{1}$ & $0$ & $1$ & $-$ \\
    \hline
    $S$ & $\mathbf{1}$ & $0$ & $1$ & $-$ \\
    \hline
  \end{tabular}
  \caption{Particle content of a natural Dirac model~\cite{CentellesChulia:2020dfh}. $\nu_R$ and $N_{L,R}$ are fermions while $S$ is a scalar. All fields are color singlets.
    \label{tab:dirac}
  }
  \end{table}

Before concentrating on Majorana neutrinos, let us say a few words about the Dirac alternative. This option is less popular, mainly because it is often restricted to the simplest Dirac neutrino mass model. In this case, one simply replicates what is done for the other SM fermions: (i) add right-handed neutrinos, $\nu_R$; (ii) write Yukawa couplings with the Higgs doublet, $Y_\nu \, \overline{\ell_L} \, \widetilde \Phi \, \nu_R$; and (iii) generate Dirac neutrino masses once the Higgs acquires a non-zero vacuum expectation value (VEV). With $Y_\nu \sim 10^{-12}$, one obtains neutrino masses in the correct ballpark. However, this simple model faces two issues:
\begin{itemize}
\item The required Yukawa couplings are very small, making the setup virtually untestable and exacerbating the flavor puzzle.
\item There is no fundamental reason not to include a Majorana mass for the right-handed neutrinos: $M_R \, \overline{\nu_R^c} \nu_R$.
\end{itemize}
For these reasons, this model is generally regarded as less attractive. However, there are natural models with Dirac neutrinos that do not suffer from these issues and have sizable Yukawa couplings and a potentially rich phenomenology. An example is given in~\cite{CentellesChulia:2020dfh}.~\footnote{See also~\cite{Roncadelli:1983ty,Roy:1984an} for pioneering work on Dirac neutrinos.} The SM particle content is extended with the fields in Tab.~\ref{tab:dirac} and the SM symmetry with a new global $U(1)_L$ group and a discrete $\mathbb{Z}_2$ parity. With these ingredients, one can write the Yukawa Lagrangian
\begin{equation}
  \mathcal{L}_{\rm D}^\nu = y_N \, \overline{\ell_L} \widetilde \Phi N_R + y_R \overline{N_L} \nu_R S + M_N \overline{N_L} N_R + \, \text{h.c.} \, .
\end{equation}
After electroweak symmetry breaking one finds the Dirac mass term
\begin{equation}
\left( \begin{array}{cc} \overline{\nu_{L}} & \overline{N_{L}} \end{array} \right)
\left( \begin{array}{cc} 0 & \ \ m_L \\ m_R & \ \ M_N \end{array} \right)
\left( \begin{array}{c} \nu_{R} \\ N_{R} \end{array} \right) + {\rm h.c.} \, ,
\end{equation}
with $m_L = y_N \langle \Phi \rangle$ and $m_R = y_R \langle S \rangle$. If $m_L \ll m_R \ll M_N$, this leads to the light neutrino masses
\begin{equation}
  m_\nu \sim m_L \, \frac{m_R}{M_N} \, ,
\end{equation}
which are naturally small thanks to the assumed hierarchy, even if the involved Yukawa couplings are sizable. This is a \textit{Dirac seesaw mechanism} at work.

\subsection{Majorana neutrinos}
\label{subsec:majorana}

The Majorana neutrino alternative is by far the preferred option among most theorists. In fact, neutrinos are \textit{expected} to be Majorana particles unless some symmetry forbids lepton number violation. For instance, in the examples discussed in Sec.~\ref{subsec:dirac}, this role was played by a global $U(1)_L$ symmetry. For this reason, all simple (and theoretically appealing) models are of Majorana type, with the type-I seesaw being the most popular example. These models are more economical than their Dirac counterparts, with just two degrees of freedom instead of four. They are also natural in the context of effective field theory, as pointed out long ago by Weinberg~\cite{Weinberg:1979sa}. Indeed, the only dimension-5 operator consistent with the SM gauge symmetry is the so-called Weinberg operator,
\begin{equation}
Q_{W} = \left( \widetilde \Phi^\dagger \ell \right)^T C \left( \widetilde \Phi^\dagger \ell \right) \sim \ell\ell\Phi\Phi \, ,
\end{equation}
which breaks lepton number in two units and leads to Majorana neutrino masses once the Higgs doublet gets a VEV. This has been vastly exploited by model builders, with plenty of realizations of the Weinberg operator in the literature. In Majorana neutrino mass models, one can write a generic expression for the light neutrino masses~\cite{Bonnet:2012kz}
\begin{equation} \label{eq:majorana}
m_\nu \propto \frac{\langle \Phi \rangle^2}{\Lambda} \, \times \, \epsilon \, \times \, \left( \frac{1}{16 \pi^2} \right)^n \, \times \, \left( \frac{\langle \Phi \rangle}{\Lambda} \right)^{d-5} \, .
\end{equation}
This expression helps identify the possible sources of neutrino mass suppression in Majorana models. The first factor is typical of seesaw scenarios and must necessarily appear due to gauge invariance. Here, $\Lambda$ denotes the scale at which neutrino masses are generated. Each of the remaining three factors corresponds to a possible suppression mechanism. The first arises from the approximate conservation of lepton number, broken by a small parameter $\epsilon$, as in low-scale seesaw models. The second appears in radiative models, where $n$ denotes the loop order at which neutrino masses are generated. Finally, the third factor is associated with models in which neutrino masses originate from operators of dimension $d > 5$.

\begin{table}
  \renewcommand*{\arraystretch}{1.4}
  \centering
  \begin{tabular}{|ccc|}
    \hline
    & $SU(2)_L$ & $U(1)_Y$ \\
    \hline
    \hline
    $\nu_R$ & $\mathbf{1}$ & $0$\\
    \hline
  \end{tabular}
  \caption{Particle content of the type-I seesaw. Three generations of right-handed neutrinos are normally assumed, all singlets under the SM gauge symmetry (including $SU(3)_c$).
    \label{tab:seesaw}
  }
\end{table}

The most popular Majorana model is the type-I seesaw~\cite{Minkowski:1977sc}. This framework extends the SM particle content by adding right-handed neutrino singlets, typically in three generations, with the quantum numbers shown in Tab.~\ref{tab:seesaw}. One can then write the Yukawa Lagrangian
\begin{equation}
  \mathcal{L}_{\rm M}^\nu = y \, \overline{\ell_L} \widetilde \Phi \nu_R + \frac{1}{2} M_R \overline{\nu_R^c} \nu_R + \, \text{h.c.} \, .
\end{equation}
After symmetry breaking, the neutral fermion mass matrix is given by
\begin{equation}
  \mathcal M = \left(\begin{array}{c c} 
0 & m_D \\ 
m_D^T & M_R
  \end{array}\right) \, ,
\end{equation}
with $m_D = y \, \langle \Phi \rangle$. Then, assuming $m_D \ll M_R$ one finds the light neutrino mases
\begin{equation}
m_\nu = - m_D \, {M_R}^{-1} \, m_D^T  \sim \frac{\langle \Phi \rangle^2}{M_R} \, .
\end{equation}
We see that only the first factor in Eq.~\eqref{eq:majorana} operates in the case of the type-I seesaw. The smallness of neutrino masses is then associated with the presence of a large energy scale, $\Lambda \sim M_R$. Other realizations of this idea exist, all sharing this general feature. This mechanism is very economical and fits naturally within grand unified theories. However, it is not directly testable. Therefore, it is interesting to consider alternative directions.

\section{The majoron}
\label{sec:majoron}

In the following we concentrate on models with spontaneous lepton number violatoin. Therefore, we will consider Lagrangians invariant under a global $U(1)_L$ symmetry, spontaneously broken by the VEV of a scalar carrying lepton number. As we will see, the main novelty in this class of scenarios is the existence of a Goldstone boson, the majoron, usually denoted by $J$.~\footnote{It has been pointed recently that the underlying symmetry behind the majoron is not $U(1)_L$ but the typically non-anomalous $U(1)_{B-L}$~\cite{Herrero-Brocal:2026nmc}. In this paper, however, we will stick to the traditional approach in terms of just lepton number. The results discussed here are not affected by this detail.}

\begin{table}
  \renewcommand*{\arraystretch}{1.4}
  \centering
  \begin{tabular}{|cccc|}
    \hline
    & $SU(2)_L$ & $U(1)_Y$ & $U(1)_L$ \\
    \hline
    \hline
    $\nu_R$ & $\mathbf{1}$ & $0$ &  $1$ \\
    $\sigma$ & $\mathbf{1}$ & $0$ &  $-2$ \\
    \hline
  \end{tabular}
  \caption{Particle content of the type-I seesaw with spontanous global $U(1)_L$ violation.
    \label{tab:seesawSLV}
  }
\end{table}

It is actually straightforward to build a variant of the type-I seesaw with spontaneous lepton number violation. One just needs to add the fields in Tab.~\ref{tab:seesawSLV} to the SM particle content. Here, $\sigma$ is a new scalar singlet with non-zero lepton number. With this addition, one can write the terms
\begin{equation}
  \mathcal{L}_{\rm I}= y \, \overline{\ell_L} \, \nu_R \, \Phi + \frac{1}{2} f \sigma \, \overline{\nu_R^c} \, \nu_R + \text{h.c.} \, .
\end{equation}
Then, when $\sigma$ gets a VEV, $\langle \sigma \rangle \neq 0$, a Majorana mass term for the right-handed neutrinos is induced, $f \langle \sigma \rangle \, \overline{\nu_R^c} \, \nu_R = M_R \, \overline{\nu_R^c} \, \nu_R$. This leads to the type-I seesaw mechanism, as discussed above. The difference with respect to the usual scenario with explicit lepton number violation is given by Goldstone's theorem. The spontaneous breaking of the continuous global $U(1)_L$ symmetry gives rise to a massless Goldstone boson, the majoron~\cite{Chikashige:1980qk,Chikashige:1980ui}.~\footnote{The majoron can be regarded as a particularly well-motivated axion-like particle. It is naturally leptophilic and has lepton flavor violating signatures~\cite{Cornella:2019uxs,Escribano:2020wua,Bauer:2021mvw}.}

In fact, lepton number only has a clear meaning (i.e., is a well-defined quantum number for model building) if its breaking is spontaneous. For instance, as explicitly shown in~\cite{CentellesChulia:2024uzv}, all singlet fermion extensions of the SM (inverse seesaw, linear seesaw, \dots) with explicit lepton number violation can be understood as equivalent to a type-I seesaw with a specific number of singlet generations and particular matrix textures. One can establish a dictionary that maps the Lagrangian of any of these models onto that of a general type-I seesaw. In the case of the inverse seesaw~\cite{Mohapatra:1986bd}, this is achieved by grouping the singlet fermions $N$ and $S$ (the two types of singlets in the inverse seesaw) into a general $\nu_R$ multiplet, with specific textures in the Yukawa Lagrangian: a vanishing $\ell_L - S$ Yukawa term and a vanishing $N - N$ Majorana mass term.

This is no longer true if the breaking of lepton number is spontaneous. For instance, in the scenario that we just described one cannot group the $N$ and $S$ singlets in a common multiplet simply because it is forbidden by $U(1)_L$ conservation (they carry different lepton numbers)~\cite{CentellesChulia:2020dfh}. And this has physical consequences. As we proceed to show now, the assignment of $U(1)_L$ charges is crucial: two models leading to the same neutrino mass mechanism may have very different phenomenological predictions. The majoron, the Goldstone boson resulting precisely from the breaking of the lepton number symmetry, is the key.

\begin{table}
  \renewcommand*{\arraystretch}{1.4}
  \centering
  \begin{tabular}{|ccccccc|}
    \hline
    & $\ell_L$ & $e_R$ & $\Phi$ & $N$ & $S$ & $\sigma$ \\
    \hline
    \hline
    Model 1: ``canonical'' & 1 & 1 & 0 & 1 & -1 & -2 \\
    Model 2: ``enhanced'' & 1 & 1 & 0 & 1 & 0 & -1 \\
    \hline
  \end{tabular}
  \caption{$U(1)_L$ charge assignment in two variants of the inverse seesaw with spontaneous lepton number violation.
    \label{tab:inverse}
  }
\end{table}

In order to illustrate this point, let us consider two variants of the inverse seesaw with spontaneous lepton number violation~\cite{CentellesChulia:2020dfh}. In addition to the SM fields, both models feature two types of fermion singlets, $N$ and $S$, as well as the $\sigma$ scalar singlet. Therefore, they have exactly the same particle content and only differ in the their lepton number assignments, as shown in Tab.~\ref{tab:inverse}.

\textbf{Model 1} (which we refer to as ``canonical'') follows the usual choice, assigning opposite lepton numbers to the $N$ and $S$ fermions, namely $1$ and $-1$, respectively. This requires $\sigma$ to carry lepton number $-2$, allowing for the Yukawa terms
  \begin{equation}
    -\mathcal{L}_1 = y_N \, \overline{\ell_L} \tilde{\Phi} N + m_R \, \bar{N}^c S + \frac{1}{2} \lambda_N \, \sigma \, \bar{N}^c N + \frac{1}{2} \lambda_S \, \sigma^* \, \bar{S}^c S + \text{h.c.} \, .
  \end{equation}
  After electroweak and $U(1)_L$ symmetry breaking, this leads to the neutral fermion mass matrix
  \begin{equation} \label{eq:minv1}
    \mathcal{M}_1 = \begin{pmatrix}
0 & y_N \, \Phi & 0 \\
y_N^T \, \Phi & \lambda_N \, \sigma & m_R \\
0 & m_R^T & \lambda_S \, \sigma^*
    \end{pmatrix} \, ,
  \end{equation}
  where the scalar fields (intead of their VEVs) have been written for the sake of clarity. The matrix in Eq.~\eqref{eq:minv1} has the structure characteristic of the inverse seesaw. In fact, if the hierarchy $\lambda_S \, v_\sigma \ll y_N \, v  \ll m_R$ is satified, an inverse seesaw mechanism operates, leading to light neutrino masses $m_\nu  \sim y_N^2 \, \lambda_S \, \frac{v^2 v_\sigma}{m_R^2}$.

\textbf{Model 2} (which we refer to as ``enhanced'') makes an alternative choice. In this case, $S$ has vanishing lepton number, which allows for a bare Majorana mass term in the Lagrangian. The Yukawa terms are now given by
  \begin{equation}
    -\mathcal{L}_2 = y_N \, \overline{\ell_L} \tilde{\Phi} N + \lambda \, \sigma \, \bar{N}^c S + \frac{m_S}{2} \, \bar{S}^c S + \text{h.c.} \, .
  \end{equation}
  They lead to the neutral fermion mass matrix
  \begin{equation} \label{eq:minv2}
  \mathcal{M}_2 = \begin{pmatrix}
0 & y_N \, \Phi & 0 \\
y_N^T \, \Phi & 0 & \lambda \, \sigma \\
0 & \lambda^T \sigma & m_S
  \end{pmatrix} \, ,
\end{equation}
which again has the usual structure obtained in the inverse seesaw. In this case, the hierarchy $m_S \ll y_N \, v  \ll \lambda \, v_\sigma$ leads to $m_\nu  \sim \frac{y_N^2}{\lambda^2} \, \frac{v^2 m_S}{v_\sigma^2}$.

Both variants lead to an inverse seesaw mechanism and feature a majoron, due to the spontaneous breaking of $U(1)_L$. At first sight, they may appear to be fundamentally equivalent, with similar phenomenological predictions. However, this is not the case, as the majoron coupling to charged leptons is dramatically different in the two scenarios.

The majoron coupling to charged leptons arises at the 1-loop level in many models, including those considered here. It was first computed for the type-I seesaw in~\cite{Chikashige:1980ui} (see also \cite{Pilaftsis:1993af,Heeck:2019guh}), and more recently in \textit{any} majoron model in~\cite{Herrero-Brocal:2023czw}. Applying the results of this latter work to the two inverse seesaw variants considered here leads to very different predictions for the majoron coupling to charged leptons, $g_{J \ell \ell}$. In the canonical model, one finds
$g_{J \ell \ell} \sim \frac{y_N^2 \lambda_S^2}{96 \pi^2} \frac{M_\ell v_\sigma}{m_R^2} \sim \frac{\lambda_S}{96 \pi^2} \frac{M_\ell m_\nu}{v^2}$, whereas in the enhanced model one obtains $g_{J \ell \ell} \sim \frac{y_N^2}{16 \pi^2} \frac{M_\ell}{v_\sigma}$. While the former is suppressed by neutrino masses, the latter is not. The origin of this difference is easy to understand. In both models, the majoron can be identified with the imaginary component of the singlet $\sigma$. In the canonical model, $\sigma$ couples to leptons via the $\lambda_S \, \sigma^* \, \bar{S}^c S$ Yukawa term, which is assumed to generate the smallest entry in the neutral fermion mass matrix. In contrast, in the enhanced model, $\sigma$ couples to leptons through the unsuppressed $\lambda \, \sigma \, \bar{N}^c S$ Yukawa interaction. This simple observation explains why the majoron coupling to charged leptons is expected to be tiny in the canonical model, while it can be sizable in the enhanced scenario.

\begin{figure}[t!]
\centering
\includegraphics[height=6.2cm]{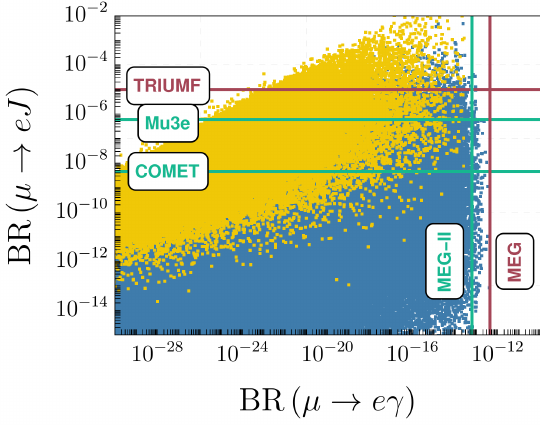}
\caption{Comparison between BR($\mu \to e \, J$) and BR($\mu \to e \, \gamma$) in the enhanced inverse seesaw model. The orange points have Yukawa couplings $y_N, \lambda \sim \mathcal{O}(1)$, while the blue points allow for Yukawa couplings in the range $y_N, \lambda \sim \mathcal{O}(10^{-3}-1)$.
\label{fig:enhanced}}
\end{figure}

This has clear phenomenological implications. In the canonical model, processes such as $\mu \to e \, J$ have negligibly small rates and are essentially irrelevant for experimental searches. In contrast, in the enhanced model they can become the dominant flavor-violating observables, as illustrated in Fig.~\ref{fig:enhanced}. This figure shows numerical results for the branching ratios of the exotic $\mu \to e \, J$ and the more conventional $\mu \to e \, \gamma$ decays. Most points satisfy BR($\mu \to e \, J$) $>$ BR($\mu \to e \, \gamma$), with the latter lying below the reach of future experiments in practically the entire parameter space. The clear conclusion of this analysis is that $\mu \to e \, J$ provides a much more sensitive probe of the model than $\mu \to e \, \gamma$.~\footnote{Cosmological probes can actually be more powerful for the case of canonical models~\cite{CentellesChulia:2025eck}.}

\section{Final remarks}
\label{sec:final}

There are many neutrino mass models. I ended up focusing on models with spontaneous lepton number violation featuring a majoron, but many more well-motivated scenarios exist. They normally have specific phenomenological predictions that allow us to search for them and, if a signal is found, distinguish among them. In fact, neutrino model building is a very active field. Many relevant subjects that are currently being explored were not covered in this talk. For instance, flavor models, modular symmetries or radiative models, just to mention a few topics that have become popular in recent years.

\providecommand{\href}[2]{#2}\begingroup\raggedright\endgroup
 
\end{document}